\documentclass[preprint,12pt]{elsarticle}
\usepackage{epsfig}
\usepackage{amssymb}
\usepackage{amsmath}
\newfont{\logo}{logo10}

\newcommand{\bea}{\begin{eqnarray}}
\newcommand{\eea}{\end{eqnarray}}
\newcommand{\bes}{\begin{subequations}}
\newcommand{\ees}{\end{subequations}}
\newcommand{\ds}{\displaystyle}

\begin{document}
\begin{frontmatter}
\title{Explicit construction of single input - single output logic gates from three soliton solution of Manakov system}

\author[label2]{M. Vijayajayanthi \corref{cor1}}
\author[rvt2]{T. Kanna}
\author[label3]{M. Lakshmanan}
\author[label2]{K. Murali}
\cortext[cor1]{Corresponding author:vijayajayanthi.cnld@gmail.com}
\address[label2]{Department of Physics, Anna University, Chennai--600 025, India}
\address[rvt2]{PG and Research Department of Physics,
 \\
Bishop Heber College,Tiruchirapalli--620 017, India}
\address[label3]{Centre for Nonlinear Dynamics, School of Physics, Bharathidasan University, Tiruchirapalli--620 024, India}
\begin{abstract}
We construct single input logic gates using the energy sharing collisions of a minimal number of (three) bright optical solitons associated with the three soliton solution of the integrable Manakov system.  As computation requires state changes to represent binary logic, here we make use of the state change of a particular soliton during its sequential collision with other two solitons for constructing single input gates.   As a consequence, we clearly demonstrate the construction of various one-input logic gates such as COPY gate, NOT gate and ONE gate using energy sharing three soliton collision of Manakov system.  This type of realization of logic gates just from a three soliton collision (pair-wise interaction) is clearly distinct from the earlier studies which require separate collisions of four solitons.
\end{abstract}


\end{frontmatter}

      

\section{Introduction}
The recent developments in optical computing, quantum computing, computing via chaos suggest that light can be used to execute logical operations instead of discrete electronic components used in the present day computer \cite{book}.  In this work, we focus on collision based computation involving energy sharing collisions of solitons in nonlinear media.  Here, computation occurs by the pair-wise collisions of solitons, where each soliton bears a finite state value before collision, and state transformations occur at the time of collisions between solitons. 

Collision based computation can be realized in several physical and chemical systems such as cellular automata \cite{sapin}, fiber couplers \cite{coupler}, Josephson junction \cite{josephson}, etc.  This collision based computing originally introduced in conservative computation such as a billiard ball model \cite{fredkin} and its cellular automaton analogues \cite{ada}, presents a novel approach of computation with mobile physical objects (e.g., billiard balls, chemical particles, self-localized patterns on cellular automata, and so on).  The well-established problems of emergent computation and universality in cellular automata have been tackled by a number of researchers in the past 35 years \cite{banks,margolus,lindgren,ada2} and still remains as an active area of theoretical computer science and nonlinear science.  The best-known universal automaton is the Game of Life \cite{gardner}.  It was shown to be universal by Conway in Ref. \cite{conway} who simulated logic gates by the Game of Life.  An evolutionary algorithm searching for collision-based computing in cellular automata has been presented in Ref. \cite{sapin}.  Here, the AND gate has been simulated by the Game of Life.   Also Adamatzky et al.  demonstrated exact implementation of basic logical operations with signals in Belousov-Zhabotinsky medium \cite{ada_BZ} and the experimental realization of logic gates has been presented in Ref. \cite{ada_BZe}.  On the other hand, in fiber couplers a full set of logic functions including AND, NAND, XOR, NOT and OR gates are numerically demonstrated using two-core and three-core fiber coupler switches operating in the continuous wave regime \cite{coupler}.  Especially, it has been shown that the logic gates AND, OR and XOR can be constructed from an asymmetric two-core fiber coupler and in the symmetrical three core fiber coupler NAND, AND, OR, XOR and NOT logical gates can be realized.  Likewise, using Josephson junction, collision based (fusion) computing has been performed in Ref. \cite{josephson}.

Using light field as carrier of information in modern day computers, which now employ electrons, has several advantages like faster speeds, smaller computers and less heat dissipation. To be specific, light creates virtually no heat when it travels while the electric current used in present day computers radiates a lot of heat.  Additionally, light has the ability to pass through other beams of light. Two laser beams (pulses) can cross each other whereas electric currents cannot do this and the present computers are designed such that they never admit cross paths.  Since the beam of light can cross each other, less space is required.  This would result in smaller computers.  Too many transistors used in modern day computers will also slow down the processor speed and metallic wires limit the speed of transmission whereas in a single light path, several data sets can be transmitted parallely at the same time using different wavelengths/polarizations.  The higher parallelism and the faster velocity of light allow extreme processing speeds. These important characteristics of light suggest us to look for optical computing. In the pioneering work, Jakubowski, Steiglitz and Squier  (JSS)\cite {jaku} designed sequences of solitons operating on other sequences of solitons that effect logic operations, including controlled NOT gate. In this paper we have shown the explicit construction of one-input logic gates using three soliton collisions.   

For this purpose, we consider the pico-second pulse propagation in a lossless strongly birefringent Kerr-type optical fiber \cite{bis1,bis2,bis3} with local and instantaneous response in the anomalous dispersion regime which is governed by the following celebrated Manakov system \cite{kivshar}:
\begin{subequations} 
\bea
iq_{1,z}+q_{1,tt}+2\Big(|q_1|^2+|q_2|^2 \Big)q_1=0,\\
iq_{2,z}+q_{2,tt}+2\Big(|q_1|^2+|q_2|^2 \Big)q_2=0,
\eea
\label{manakov}
\end{subequations}
where $q_1$ and $q_2$ are the complex amplitudes of the first and second components, the subscripts $z$ and $t$ represent the partial derivatives of the normalized distance along the fibre and the retarded time, respectively.  Here, solitons are used to carry the information inside the nonlinear medium and computation occurs when these solitons collide.  Manakov solitons have been observed experimentally in Ref. \cite{expt}.  These Manakov solitons undergo fascinating energy sharing collision as well as standard elastic collision .   Radhakrishnan, Lakshmanan and Hietarinta \cite{hie} have obtained  the two-soliton solution of the integrable Manakov system and revealed that solitons in this system exhibit intriguing shape changing or energy sharing collisions which subsequently have been well studied in Refs. \cite{kannaprl,kannapre,pramana} and various types of energy sharing collisions have been observed in different multicomponent systems in \cite{kannapre,coupling,r_pre,r_pre2,r_pre3} .   However, computation requires state changes to represent binary logic and the energy sharing collision properties of Manakov solitons suggest their feasibility for computation.   A salient feature of this kind of computation is that it performs conservative logic operations as the collisions preserve total energy of the Manakov system irrespective of their state change.  Collision dynamics of energy sharing solitons in Manakov system and its application in computing are studied in detail in  Refs. \cite{kannaprl,kannapre,pramana}.  Below, we first review this exciting energy sharing collision and then discuss the principle and our construction procedure of one-input logic gates.
\section{Brief review of developments of soliton collision based optical logic gates construction}
The Manakov system (\ref{manakov}) describes the propagation of an intense electromagnetic wave in a two mode /birefringent fibre as mentioned in the introduction.  Here, interaction between the field components results in intensity dependent nonlinear cross-coupling terms.  In 1973, Manakov \cite{man} solved the set of coupled nonlinear evolution equations (\ref{manakov}) using the inverse scattering transform method and obtained multisoliton solutions.  Later, in Ref. \cite{hie} fascinating collision properties of bright solitons in the Manakov system have been revealed. Also, here the polarization parameters bring an additional freedom so that there occurs an amplitude/intensity redistribution among the colliding solitons.  As a consequence of this, a particular soliton in a given component can enhance its intensity along with suppression in the other component.  There will also be commensurate changes in the other soliton.  Thus in this interesting collision process, solitons in a given component exchange energy in order to conserve the energy in that component.  Additionally, the solitons in different components also exchange energy so that the total intensity is also conserved.  JSS \cite{jaku} found that this energy sharing collision can be profitably used for performing nontrivial information transformation.  In Ref. \cite{jaku} the state change undergone by each colliding soliton was expressed as a linear fractional transformation (LFT).  This transformation depends on the total energy and velocity of the solitons which are invariant under collision.  JSS designed sequences of solitons operating on other sequences of solitons that effect logic operations, including controlled NOT gate.  In this operation, both data and logic operators have the self-restoring and re-usability features of digital logic circuits.  Also, numerical simulation of an energy switching NOT processor was implemented in the Manakov system.  It was suggested that it might be possible to use these light-light interactions to do general computation in a bulk medium, without interconnecting discrete components.

In 2000, Steiglitz \cite{steig} extended the work to the Manakov (1+1) dimensional spatial solitons for performing arbitrary computation in a homogeneous medium with beams entering only at one boundary.  Here both the dimensions are spatial.  For computational purpose, separate collisions of {\it four solitons} (three down moving vertical solitons and one left moving horizontal soliton known as actuator soliton) were considered for computing one-input gates.  The down moving vertical solitons collide with the fixed actuator soliton and therefore the use of three collisions and a fixed actuator makes more flexible gates possible.  The COPY gate, FANOUT gate , NOT gate and the universal NAND gate were designed theoretically in Ref. \cite{steig}.  Rand et al. \cite{rand} have shown the signal standardization in collision based soliton computing.  It is completely based on bistable configuration of Manakov solitons.  Kanna and Lakshmanan \cite{kannapre,kannaprl} have shown systematically how the LFT follows from the two soliton collision, and generalized the results to multicomponent case which has the advantage of realizing multistate logic rather than binary logic. There itself a theoretical study has been presented to view the four soliton collision as one-input gate and the corresponding mathematical conditions have been derived.  It is of natural interest to look for less number of soliton collisions for constructing logic gates by incorporating new ideas and modifying the earlier suggestions for further development of this technologically important topic of research. {\it For this purpose, in this paper we focus to construct the one-input gates such as COPY, NOT and ONE gate using only by the collisions of three solitons.  This type of realization of logic gates just from a three soliton collision (pair-wise interaction) is clearly distinct from the earlier studies which require separate collisions of four solitons.}  As computation requires state changes, below we define suitable complex state vector to represent the state of a soliton as was done in Ref. \cite{kannapre}.
\section{Soliton characterization of Manakov system}
Let us first consider the one soliton solution obtained in Ref. \cite{kannapre} using the Hirota's method.  One can write down the explicit one-soliton solution as
\begin{eqnarray}
\left(
\begin{array}{c}
q_1\\
q_2 
\end{array}
\right)
 = 
\left(
\begin{array}{c}
\alpha_1^{(1)}\\
\alpha_1^{(2)}
\end{array}
\right)\frac{e^{\eta_1}}{1+e^{\eta_1+\eta_1^*+R}}\;\;
 = 
\left(
\begin{array}{c}
A_1 \\
A_2
\end{array}
\right)
\frac{k_{1R}e^{i\eta_{1I}}}{\mbox{cosh}\,(\eta_{1R}+\frac{R}{2})},
\end{eqnarray}                                                                                                          
where $\eta_1=k_1(t+ik_1z)=\eta_{1R}+i \eta_{1I}$, $A_j=\frac{\alpha_1^{(j)}}
{\left[\left(|\alpha_1^{(1)}|^2+|\alpha_1^{(2)}|^2\right)\right]^{1/2}}$, $j=1,2$ 
and $e^R=\frac{\left(|\alpha_1^{(1)}|^2+|\alpha_1^{(2)}|^2\right)}{(k_1+k_1^*)^2}
$.
Note that the above one-soliton solution is characterized by three arbitrary complex 
parameters $\alpha_1^{(1)}$,  $\alpha_1^{(2)}$ and $k_1$. Here the 
amplitude of the soliton in the first and second components (modes) are given 
by $k_{1R}A_1$ and $k_{1R}A_2$, respectively, subject to the condition 
$|A_1|^2+|A_2|^2=1$, while the soliton velocity in both the modes  
is given by $2k_{1I}$. Here $k_{1R}$ and $k_{1I}$ represent the real and
imaginary parts of the complex parameter $k_1$. The quantity 
$\frac{R}{2k_{1R}}=\frac{1}{2k_{1R}}$ $\mbox{ln}$$\left[
\frac{\left(|\alpha_1^{(1)}|^2+|\alpha_1^{(2)}|^2\right)}{(k_1+k_1^*)^2}
\right]
$ denotes the position of the soliton.

In a similar fashion, one can characterize the two-soliton and three-soliton solutions respectively by six ($\alpha_i^{(j)}$'s and $k_i$'s, $i,j=1,2$) and nine ($\alpha_i^{(j)}$'s and $k_i$'s, $i=1,2,3$, $j=1,2$) arbitrary complex parameters. For the explicit expression of two and three soliton solutions, one can refer to \cite{kannapre}.
\section{Theory of construction of one-input$-$one-output gates}
\subsection{State of a soliton}
\begin{figure}[h]
\begin{center}
\epsfig{figure=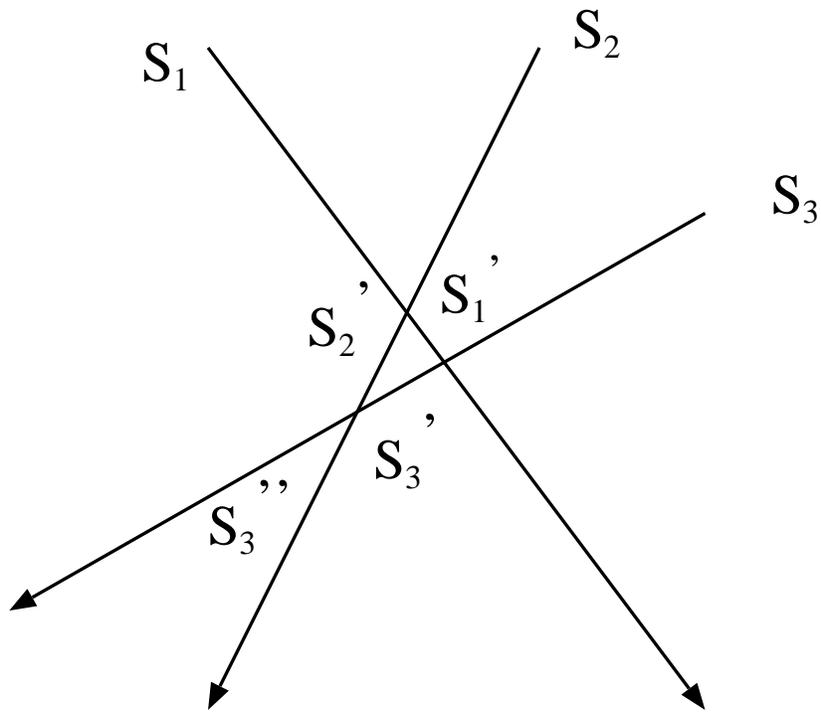, width=0.8 \columnwidth}
\caption{Collision picture of solitons $S_1$, $S_2$ and $S_3$.}\label{3_collision}
\end{center}
\end{figure}

 We consider the collision of three solitons. This collision is completely described by the three-soliton solution given in Appendix, in Gram determinant form for simplicity. Here, we assume that $k_{jR}>0$; $k_{1I}>k_{2I}>k_{3I}$, without loss of generality, where the suffices $R$ and $I$ in $k_j$ denote the real and imaginary parts.  According to our assumption, the first collision occurs between solitons $S_1$ and $S_2$, then the resulting soliton, say $S_1^{\prime}$, after the first collision is allowed to collide with the third soliton $S_3$.  Finally, the resulting solitons $S_2^{\prime}$ and $S_3^{\prime}$ collide with each other. Schematic representation of the above discussed three-soliton collision sequence is given in Fig. \ref{3_collision}.  One can look into Refs. \cite{kannapre,pramana} for a clear picture of this collision process, where detailed asymptotic analysis of the above mentioned three soliton solution has been presented.  Let the state of any soliton be represented by a complex state vector ($\rho$).  If we consider the single soliton solution spread up in two components then $\rho$ is given by the ratio of the two components 
\bea
\rho=\frac{q_1}{q_2}.
\eea
 Here in the three soliton collision, we deal with the asymptotic states of the three colliding solitons.
The corresponding asymptotic states of the $j$th ($j=1,2,3$) soliton are defined below:
\bea
\rho^{j\pm}=\frac{q_1^j(z\rightarrow\pm\infty)}{q_2^j(z\rightarrow\pm\infty)}=\frac{A_1^{j\pm}}{A_2^{j\pm}}, \quad j=1,2,3.
\label{rhos}
\eea
where $A_{1,2}^{j\pm}$ are the polarization components (1,2) of the solitons (1,2 or 3) in a three soliton collision process.  In the expression (\ref{rhos}), suffix denotes the components, $+(-)$ denote after (before) collision and superscript $j$ represents the soliton number.
For completeness, we present the expressions of $A_{1,2}^{2-}$ and $A_{1,2}^{3+}$ in the Appendix.  
 From the asymptotic analysis, the state of soliton $S_2$ before interaction is represented by
\bea
\rho^{2-}=\frac{A_1^{2-}}{A_2^{2-}}=\frac{N_1^{2-}}{N_2^{2-}}=\frac{\alpha_1^{(1)} \kappa_{21}-\alpha_2^{(1)} \kappa_{11}}{\alpha_1^{(2)} \kappa_{11}-\alpha_2^{(2)} \kappa_{21}},\label{s2rho}
\eea
Similarly, the state of soliton $S_3$ after interaction is evaluated as

\bea
\rho^{3+}=\frac{A_1^{3+}}{A_2^{3+}}=\frac{\alpha_3^{(1)}}{\alpha_3^{(2)}}. \label{s3rho}
\eea  
\subsection {Idea behind the construction}
In general, the collisions of solitons inside the nonlinear medium implement computation where the data inputs are encoded in the states of the incoming solitons and the output of the system is encoded in the state of a particular outgoing soliton after collision.  It is not necessary for every soliton involved in the collisions to carry the input or the output data; some soliton may be needed to supply non-data inputs (to induce energy sharing collision).  Here, as an example, soliton $S_1$ is just used to induce the shape changing/energy sharing interaction among the three solitons. The input of a particular logic gate is assigned to soliton $S_2$ (which can be viewed as an input port) before interaction and the favorable output is drawn from a particular outgoing soliton (say $S_3^{\prime\prime}$) resulting after two  consecutive interactions of $S_3$ with $S_1$ and $S_2^{\prime}$, respectively. 

Having explained the physical set up, now we propose a simple method of constructing one-input gates like COPY, NOT, and ONE gates using the above equations (\ref{s2rho}) and (\ref{s3rho}). In the context of collision based optical computing, $``0"$ and $``1"$ states are represented by the ratio of the polarization components of the solitons.  Before interaction, if that ratio for the soliton $S_2$ is greater than a threshold value (say 1) then we denote the corresponding input state as $``1"$ state.  In a similar fashion, the $``0"$ state is attained when the ratio (\ref{s2rho}) is less than that threshold value.  Note that the threshold value can be complex.  However, here for convenience we assume it to be real.  For convenience, we choose the parameters of soliton 1 ($S_1$) as $\alpha_1^{(1)}=\alpha_1^{(2)}=1$.  Using this, in Eq.(\ref{s2rho}) we get
\bea
|\rho^{2-}|^2=\frac
{A A^*|Z|^2+AB^*Z+BA^*Z^*+BB^*}{BB^*|Z|^2+AB^*Z^*+BA^*Z+A A^*},\label{rho2}
\eea
where $A=k_1-2k_2-k_1^*$, $B=k_1+k_1^*$ and $Z=\frac{\alpha_2^{(1)}}{\alpha_2^{(2)}}$.  From a physical point of view intensities are measurable quantities. Ultimately, in terms of intensities the conditions for achieving $``1"$ and $``0"$ states are given respectively by $|\rho^{2-}|^2>1$ and $|\rho^{2-}|^2<1$.  In Eq. (\ref{rho2}), we also consider the parameter $Z$ as real so that $\alpha_2^{(1)}$ and $\alpha_2^{(2)}$ are real.  For achieving the $``1"$ state ($|\rho^{2-}|^2>1$), we obtain a constraint condition on the parameter $Z$ as $\frac{\alpha_2^{(1)}}{\alpha_2^{(2)}}>1$ from Eq. (\ref{rho2}) and  $\frac{\alpha_2^{(1)}}{\alpha_2^{(2)}}<1$ for achieving the $``0"$ state at the input.  However, it is not difficult to extend the analysis for complex $Z$.  If $Z$ is complex, then the conditions for achieving $``1"$ and $``0"$ states at the input are given by the following equations, respectively.
\bea
(AA^*-BB^*) (1-|Z|^2)>DZ+D^*Z^*,\\
(AA^*-BB^*) (1-|Z|^2)<DZ+D^*Z^*,
\eea
where $D=A B^*-B A^*$.
\section {Demonstration of construction of one-input logic gates using three soliton collisions}
\subsection{COPY gate}
\begin{figure}
\begin{center}
\epsfig{figure=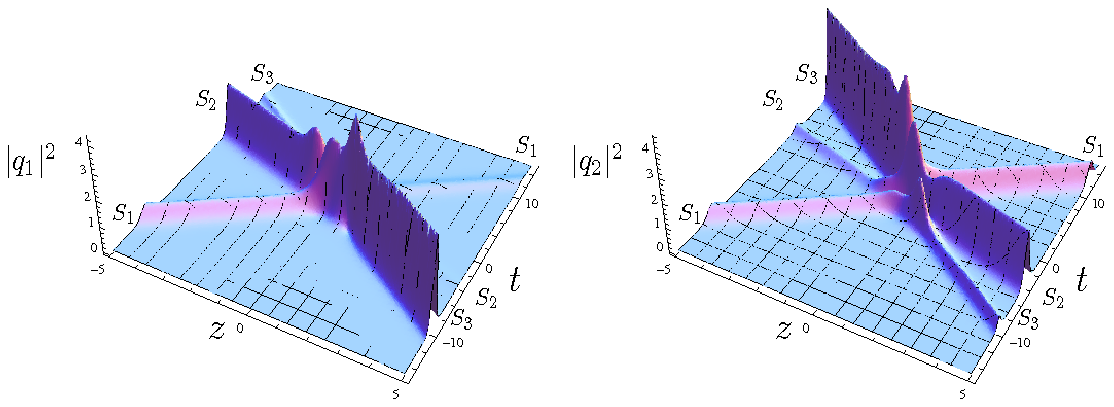, width=1.1 \columnwidth}
\caption{COPY gate: Copying $``1"$ state from the input of soliton $S_2$ to the output of soliton $S_3$ with $\alpha_2^{(1)}>\alpha_2^{(2)}$.}\label{copy_1}
\end{center}
\end{figure}

\begin{figure}
\begin{center}
\epsfig{figure=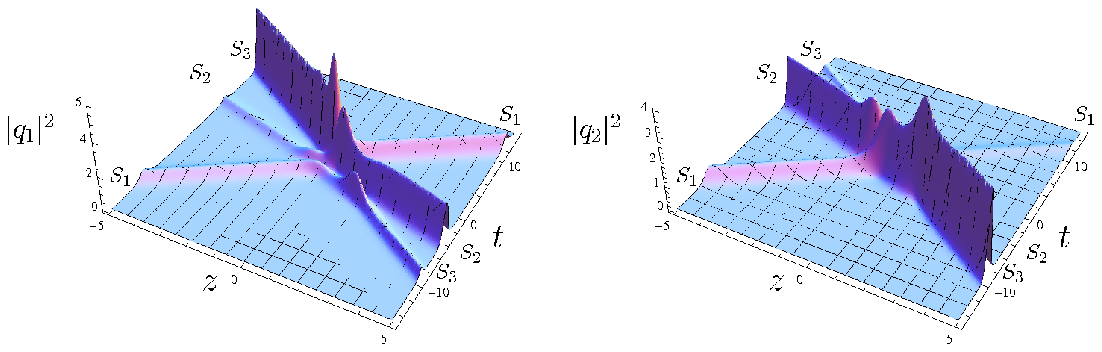, width=1.1 \columnwidth}
\caption{COPY gate: Copying $``0"$ state from the input of soliton $S_2$ to the output of soliton $S_3$ with $\alpha_2^{(1)}<\alpha_2^{(2)}$.}\label{copy_0}
\end{center}
\end{figure}

\begin{table}[!ht]
\begin{minipage}{.5\linewidth}
\caption{Truth table of COPY gate}
\centering
\begin{tabular}{|c|c|}
\hline
 Input ($S_2$) & Output ($S_3^{\prime\prime}$) \\\hline
  0& 0 \\ \hline
   1& 1 \\ \hline
\end{tabular}
\end{minipage}%
\begin{minipage}{.5\linewidth}
\centering
\caption{Intensity table of COPY gate}
\begin{tabular}{|c|c|}
\hline
 Input ($S_2$) & Output ($S_3^{\prime\prime}$) \\\hline
  0.1& 0.1 \\ \hline
   6.6& 6.6\\ \hline
\end{tabular}
\end{minipage}
\end{table}
To start with, let us consider copying the input $``1"$ state at the output using three soliton collisions.  For this purpose, the three solitons $S_1, S_2$ and $S_3$ are allowed to collide with one another in a sequence as depicted in Fig. \ref{3_collision}, where soliton $S_1$ is just used to induce the shape changing collision among the three solitons.  For convenience, again we assume the soliton $S_1$ parameters as $\alpha_1^{(1)}=\alpha_1^{(2)}=1$ and the parameter $Z$ as real.  In order to copy the $``1"$ input state of the soliton $S_2$ (before interaction) to the outgoing  soliton $S_3^{\prime\prime}$ (after interaction), we require the following conditions to be imposed on soliton $S_2$ and soliton $S_3$ parameters which are obtained from equations (\ref{s2rho}) and (\ref{s3rho}):   
\begin{subequations}
\bea 
&&\alpha_2^{(1)}>\alpha_2^{(2)},\\
&&\alpha_3^{(1)}=\frac{(k_1-k_2)\alpha_2^{(1)}-(k_2+k_1^*)\alpha_2^{(1)}+(k_1+k_1^*)\alpha_2^{(2)}}{(k_1-k_2)\alpha_2^{(2)}-(k_2+k_1^*)\alpha_2^{(2)}+(k_1+k_1^*)\alpha_2^{(1)}} \alpha_3^{(2)}.
\eea
\end{subequations}
 Figure \ref{copy_1} depicts the copying of $``1"$ state of soliton $S_2$ at the output port $S_3^{\prime\prime}$ with the parameters fixed at $k_1 = 1 +i, k_2 = 1.5 - 0.5 i, k_3 = 
 2 - i$, $ \alpha_2^{(1)} = 5, \alpha_2^{(2)} = 2$, $\alpha_3^{(2)} = 0.5 - 0.2 i$.  Theoretically, we define that $``1"$ input state is attained when the ratio of polarization components and hence the ratio of intensities of the asymptotic forms of solitons in both the components is greater than 1. This is verified by calculating the ratios of the intensities of the soliton $S_2$ well before collision (input) and the soliton $S_3$ after collision (output).  Indeed, the calculated values are $|\rho^{2-}|^2=|\rho^{3+}|^2=6.6$ which is greater than the intensity ratio threshold value 1 and ensure tat they are in $``1"$ state.  Figure \ref{copy_0} demonstrates the process of copying $``0"$ state with the same parametric choice except that $\alpha_2^{(1)}<\alpha_2^{(2)}$ and the values are $ \alpha_2^{(1)} = 2, \alpha_2^{(2)} = 5$.  The calculated values of the ratios of intensities of the solitons $S_2$ and $S_3$ are $|\rho^{2-}|^2=|\rho^{3+}|^2= 0.1$, which are less than the intensity ratio threshold value 1 and hence they are in  the $``0"$ state.  Thus, we have copied $``0"$ state of soliton $S_2$ at the input to the outgoing soliton $S_3$.  Since the Manakov system (\ref{manakov}) is dimensionless, the intensities of solitons have no units.  The truth table and the corresponding intensity table of COPY gate are given in tables I and II.
\subsection{NOT gate}
\begin{figure}
\begin{center}
\epsfig{figure=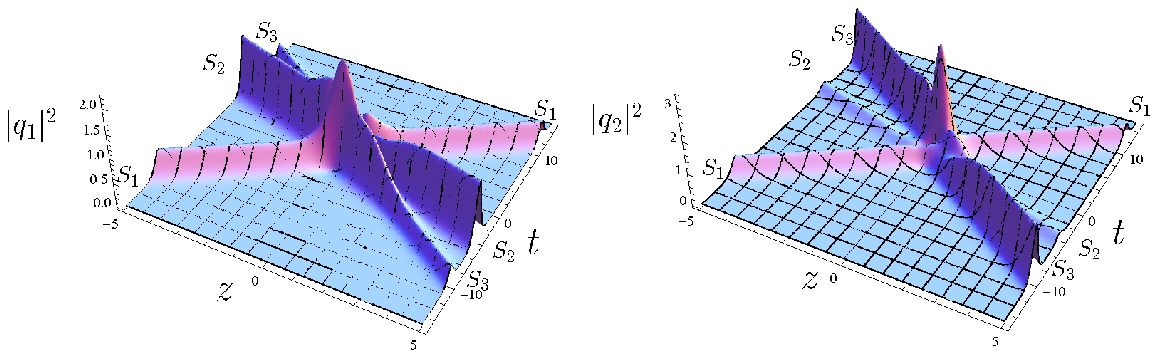, width=1.1 \columnwidth}
\caption{NOT gate: Flipping $``1"$ input state in soliton $S_2$ into $``0"$ output state in soliton $S_3$ with $\alpha_2^{(1)}>\alpha_2^{(2)}$.}\label{not_1}
\end{center}
\end{figure}

\begin{figure}
\begin{center}
\epsfig{figure=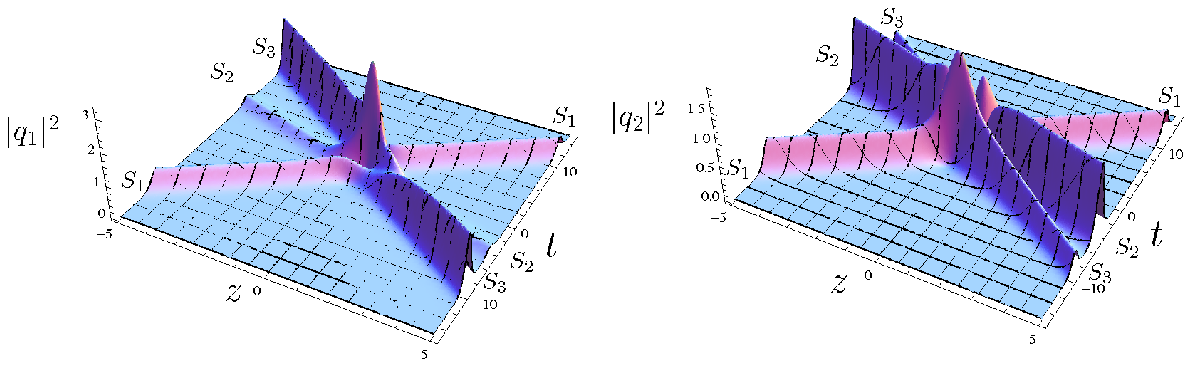, width=1.1 \columnwidth}
\caption{NOT gate: Flipping $``0"$ input state in soliton $S_2$ into $``1"$ output state in soliton $S_3$ with $\alpha_2^{(1)}<\alpha_2^{(2)}$.}\label{not_0}
\end{center}
\end{figure}

\begin{table}[!ht]
\begin{minipage}{.5\linewidth}
\caption{Truth table of NOT gate}
\centering
\begin{tabular}{|c|c|}
\hline
 Input ($S_2$) & Output ($S_3^{\prime\prime}$) \\\hline
  0& 1 \\ \hline
   1& 0 \\ \hline
\end{tabular}
\end{minipage}%
\begin{minipage}{.5\linewidth}
\centering
\caption{Intensity table of NOT gate}
\begin{tabular}{|c|c|}
\hline
 Input ($S_2$) & Output ($S_3^{\prime\prime}$) \\\hline
  0.2& 4.7 \\ \hline
   5.2& 0.2\\ \hline
\end{tabular}
\end{minipage}
\end{table}
Similar to the COPY gate, here also input is given to soliton $S_2$ before interaction and the favorable output is taken from soliton $S_3$ after interaction.  Figure \ref{not_1} shows flipping of $``1"$ input state in soliton $S_2$ before interaction into $``0"$ output state in soliton $S_3$ after interaction for the parametric choice $k_1 = 1 +i, k_2 = 1.2 - 0.5 i, k_3 =  1.4 - i$, $ \alpha_2^{(1)} =4, \alpha_2^{(2)} = 1$, $\alpha_3^{(1)} = 0.5 - 0.2 i$.  Here $\alpha_2^{(1)}$ is greater than $\alpha_2^{(2)}$, for which the input of soliton $S_2$ is in $``1"$ state. From the figure, the calculated value of ratio of intensities is 5.2, i.e. $|\rho^{2-}|^2=5.2$($\approx 5$).  In order to get the favorable output, the requirement on the $\alpha$ parameter of soliton $S_3$ is
\bea  
\alpha_3^{(2)}=\bigg[\frac{(k_1-k_2)\alpha_2^{(1)}-(k_2+k_1^*)\alpha_2^{(1)}+(k_1+k_1^*)\alpha_2^{(2)}}{(k_1-k_2)\alpha_2^{(2)}-(k_2+k_1^*)\alpha_2^{(2)}+(k_1+k_1^*)\alpha_2^{(1)}}\bigg] \alpha_3^{(1)}.  
\eea
After interaction, the intensity of soliton $S_3$ is 0.2 (i.e. $|\rho_{1,2}^{3+}|^2=0.2$) which is less than the intensity ratio threshold value 1 and hence the output is in $``0"$ state.   In a similar fashion, with the above parametric choice except that $\alpha_2^{(1)}<\alpha_2^{(2)}$, the $``0"$ input state in soliton $S_2$ is flipped into $``1"$ output state in soliton $S_3$ as shown in Fig. \ref{not_0}, where $ \alpha_2^{(1)} =1$ and $ \alpha_2^{(2)} = 4$.  The ratios of intensities of $S_2$ and $S_3$ before and after interaction are calculated as 0.2 and 4.7 (i.e, $|\rho^{2-}|^2=0.2,|\rho^{3+}|^2=4.7 (\approx 5)$), respectively, corresponding to $``0"$ input and $``1"$ output states.  The truth table and the corresponding intensity tables are also given (see tables III and IV).
\subsection {ONE gate}
\begin{figure}
\begin{center}
\epsfig{figure=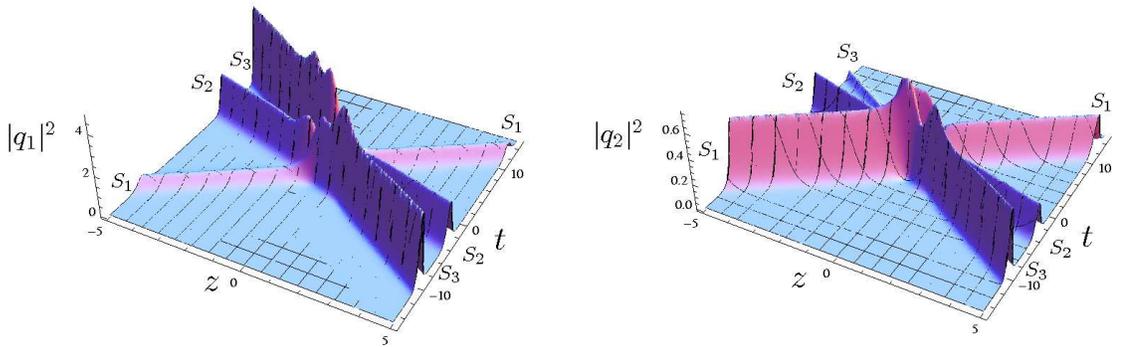, width=1.1 \columnwidth}
\caption{ONE gate: $``1"$ input and output states in solitons $S_2$ and $S_3$, respectively with $\alpha_2^{(1)}>\alpha_2^{(2)}$.}\label{one_1}
\end{center}
\end{figure}

\begin{figure}
\begin{center}
\epsfig{figure=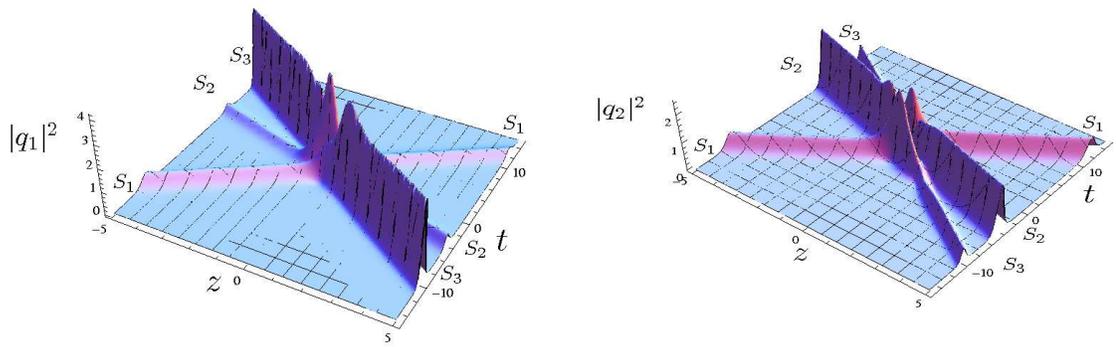, width=1.1 \columnwidth}
\caption{ONE gate: $``1"$ output state in solitons $S_3$ and $``0"$ input state in soliton $S_2$ with $\alpha_2^{(1)}<\alpha_2^{(2)}$.}\label{one_0}
\end{center}
\end{figure}

\begin{table}[!ht]
\begin{minipage}{.5\linewidth}
\caption{Truth table of ONE gate}
\centering
\begin{tabular}{|c|c|}
\hline
 Input ($S_2$) & Output ($S_3^{\prime\prime}$) \\\hline
  0& 1 \\ \hline
   1& 1 \\ \hline
\end{tabular}
\end{minipage}%
\begin{minipage}{.5\linewidth}
\centering
\caption{Intensity table of ONE gate}
\begin{tabular}{|c|c|}
\hline
 Input ($S_2$) & Output ($S_3^{\prime\prime}$) \\\hline
  0.1& 6.8 \\ \hline
   6.6& 6.6\\ \hline
\end{tabular}
\end{minipage}
\end{table}
In order to get the $``1"$ output state always irrespective of the input states, we restrict the  soliton $S_3$ parameters such that $|\alpha_3^{(1)}/\alpha_3^{(2)}|^2>1$.  If  $\alpha_2^{(1)}>\alpha_2^{(2)}$, that is $``1"$ input state is fed  into soliton $S_2$ before interaction, we get $``1"$ output state in soliton $S_3$
for the parametric choice $k_1 = 1 +i, k_2 = 1.5 - 0.5 i, k_3 =  2- i$, $ \alpha_2^{(1)} =5, \alpha_2^{(2)} = 2$, $\alpha_3^{(1)} =2.5807$, $\alpha_3^{(2)} =1$ which is shown in Fig. \ref{one_1}. This is verified by calculating the ratio of the intensities of solitons $S_2$ and $S_3$ in both the components $q_1$ and $q_2$. They are obtained as  $|\rho^{2-}|^2=|\rho^{3+}|^2= 6.6$.  Here both of the input and output solitons are in  $``1"$ state.  Figure \ref{one_0} again depicts the one gate for the above parametric choices except that $\alpha_2^{(1)}<\alpha_2^{(2)}$,for which the input state in soliton $S_2$ is $``0"$ state. The parametric values of $\alpha_2^{(1)}$ and $\alpha_2^{(2)}$ are 2 and 5, respectively.  The calculated values of the ratios of intensities of solitons $S_2$ and $S_3$ before and after interaction are $|\rho^{2-}|^2=0.1$ and $|\rho^{3+}|^2=6.8$, respectively, corresponding to $``0"$ input and $``1"$ output states.  One can refer tables V and VI for the truth table and the intensity table of ONE gate.
\section{Conclusion}
We have explicitly demonstrated the construction of various one-input logic gates, namely COPY gate, NOT gate and ONE gate, using energy sharing three soliton collisions of the Manakov system.    The principle behind the construction of such gates is to fix a threshold for the state variable and designate the states above the threshold as $``1"$ state and those below the threshold as $``0"$ state.  Here, for demonstration purpose we have chosen the threshold values to be real.  However, they can also be complex.  In a three soliton collision process, by considering a particular soliton before interaction as an input soliton, one can perform the desired logic operation on the other colliding soliton after interaction.  It is remarkable to notice that for a given gate (NOT/ONE/COPY) the numerical values of the  $``1"$ and  $``0"$ state are almost same at the input port and output port.  Here the  successive energy transfer is achieved by an activator soliton $S_1$.  Thus for the construction of one-input gate, we require only three solitons.  This type of realizing logic gates just from a three soliton expression describing pair-wise interaction of three soliton is clearly distinct from the earlier studies \cite{jaku,steig} which require separate collisions of four solitons for realizing NOT/ONE/COPY operation.  Thus our proposal of realizing multisoliton structures (exhibiting by multisoliton collisions) themselves as logic gates will have its own advantage in their experimental realization.   To our knowledge, for the first time we have explicitly explained the functioning of logic gates using three soliton solution itself.  We do believe that this study will develop a new avenue towards the experimental studies on optical computing using optical soliton collisions based on multisoliton solutions.  Using this idea, one can also construct two input logic gates including universal NOR/NAND gate with higher number of solitons.  Work is in progress in this direction.
\section*{Acknowledgement}
M. V. acknowledges the financial support from UGC - Dr. D. S. Kothari post doctoral fellowship scheme.  The work of M. L. is supported by a DST-IRPHA project.  M. L. is also supported by a DAE Ramanna Fellowship.
\appendix
\section{Three soliton solution of the Manakov system}
We can write down the three soliton solution of the Manakov system in Gram determinant form as below:
\begin{subequations}
\bea
q_s=\frac{g^{(s)}}{f}, \quad s=1,2.
\eea
where
\bea
g^{(s)}=
\left|
\begin{array}{ccccccc}
A_{11} & A_{12}& A_{13}&1&0&0& e^{\eta_1}\\
A_{21} & A_{22}& A_{23}&0&1&0& e^{\eta_2}\\
A_{31} & A_{32}& A_{33}&0&0&1& e^{\eta_3}\\
-1&0&0  & B_{11} &B_{12} &B_{13}& 0\\
0&-1&0  & B_{21} &B_{22}&B_{23} & 0\\
0&0 &-1 & B_{31} &B_{32}&B_{33} & 0\\
0&0&0& -\alpha_1^{(s)}&-\alpha_2^{(s)}&-\alpha_3^{(s)} & 0
\end{array}
\right|,
\eea
\bea
f= 
\left|
\begin{array}{cccccc}
A_{11} &A_{12} &A_{13}& 1&0&0\\
A_{21} &A_{22}&A_{23}& 0&1&0\\
A_{31} &A_{32}&A_{33}& 0&0&1\\
-1&0&0 & B_{11}&B_{12}&B_{13} \\
0&-1&0 & B_{21}&B_{22}&B_{22} \\
0&0&-1 & B_{31}&B_{32}&B_{32} \\
\end{array}
\right|,
\eea
\label{3sol}
\end{subequations}
\noindent where $A_{ij}=\ds{\frac{e^{\eta_i+\eta_j^*}}{k_i+k_j^*}}$, and $B_{ij}=\kappa_{ji}=\ds{\frac{\left(\sum_{s=1}^{N}\alpha_j^{(s)}\alpha_i^{(s)*}\right)}{(k_j+k_i^*)}}$, \;\;\;$i,j=1,2,3$.  One can refer Eq. (10) of Ref. \cite{kannapre} for the explicit expression of the above Gram determinant form of three soliton solution.
\section {Asymptotic analysis of 3-soliton solution of the Manakov system}
Considering the above three soliton solution (\ref{3sol}), 
without loss of generality, we assume that the 
quantities $k_{1R}$,
$k_{2R}$, and $k_{3R}$ are positive and  $k_{1I}>k_{2I}>k_{3I}$ .
For this condition, the variables $\eta_{iR}$'s, $i=1,2,3,$ 
for the three solitons ($S_1$, $S_2$, and $S_3$) take the following 
values asymptotically: 

(i) $\eta_{1R} \approx 0$,  $\eta_{2R} \rightarrow \pm \infty$,
$\eta_{3R} \rightarrow \pm \infty$, as $z \rightarrow \pm \infty$,

(ii) $\eta_{2R} \approx 0$,  $\eta_{1R} \rightarrow \mp \infty$,
$\eta_{3R} \rightarrow \pm \infty$, as $z \rightarrow \pm \infty$,

(iii) $\eta_{3R} \approx 0$,  $\eta_{1R} \rightarrow \mp \infty$,
$\eta_{2R} \rightarrow \mp \infty$, as $z \rightarrow \pm \infty$.

We have the following limiting forms  \cite{kannapre} of the above three-soliton solution.

\noindent\underline{(i)Before Collision (limit $ z \rightarrow -\infty$)} 

\noindent(a) \underline{\it Soliton 1}  ($\eta_{1R} \approx 0$, $\eta_{2R} \rightarrow -\infty$,
$\eta_{3R} \rightarrow -\infty$):
\begin{subequations}
\begin{eqnarray}
\left(
\begin{array}{c}
q_1\\
q_2 
\end{array}
\right)
&\approx &
\left(
\begin{array}{c}
A_1^{1-} \\
A_2^{1-}
\end{array}
\right)k_{1R}
{\mbox{sech}\,\left(\eta_{1R}+\frac{R_1}{2}\right)}e^{i\eta_{1I}} ,\\
\left(
\begin{array}{c}
A_1^{1-} \\
A_2^{1-}
\end{array}
\right)
&=&
\left(
\begin{array}{c}
\alpha_1^{(1)}\\
\alpha_1^{(2)}
\end{array}
\right) \frac{e^{\frac{-R_1}{2}}}{(k_1+k_1^*)}.
\eea
\ees
\noindent(b) \underline{\it Soliton 2}  ($\eta_{2R} \approx 0$, $\eta_{1R} \rightarrow \infty$,
$\eta_{3R} \rightarrow -\infty$):
\bes
\begin{eqnarray}
\left(
\begin{array}{c}
q_1\\
q_2 
\end{array}
\right)
&\approx&
\left(
\begin{array}{c}
A_1^{2-} \\
A_2^{2-}
\end{array}
\right)k_{2R}
{\mbox{sech}\,\left(\eta_{2R}+\frac{R_4-R_1}{2}\right)}e^{i\eta_{2I}} ,\\
\left(
\begin{array}{c}
A_1^{2-}\\
A_2^{2-}
\end{array}
\right)
&=&
\left(
\begin{array}{c}
e^{\delta_{11}}\\
e^{\delta_{12}}
\end{array}
\right)\frac{e^{-\frac{(R_1+R_4)}{2}}}{(k_2+k_2^*)}.
\eea
\ees

\noindent (c) \underline{\it Soliton 3}  ($\eta_{3R} \approx 0$, $\eta_{1R} \rightarrow \infty$,
$\eta_{2R} \rightarrow \infty$):
\bes
\begin{eqnarray}
\left(
\begin{array}{c}
q_1\\
q_2 
\end{array}
\right)
&\approx&
\left(
\begin{array}{c}
A_1^{3-} \\
A_2^{3-}
\end{array}
\right)k_{3R}
{\mbox{sech}\,\left(\eta_{3R}+\frac{R_7-R_4}{2}\right)}e^{i\eta_{3I}} ,\\
\left(
\begin{array}{c}
A_1^{3-}\\
A_2^{3-}
\end{array}
\right)
&=&
\left(
\begin{array}{c}
e^{\tau_{11}}\\
e^{\tau_{12}}
\end{array}
\right)\frac{e^{-\frac{(R_4+R_7)}{2}}}{(k_3+k_3^*)}.
\eea
\end{subequations}

\noindent \underline{(ii)After Collision (limit $ z \rightarrow +\infty$)} 

\noindent (a) \underline{\it Soliton 1}  ($\eta_{1R} \approx 0$, $\eta_{2R} \rightarrow \infty$,
$\eta_{3R} \rightarrow \infty$):
\begin{subequations}
\begin{eqnarray}
\left(
\begin{array}{c}
q_1\\
q_2 
\end{array}
\right)
&\approx&
\left(
\begin{array}{c}
A_1^{1+} \\
A_2^{1+}
\end{array}
\right)k_{1R}
{\mbox{sech}\,\left(\eta_{1R}+\frac{R_7-R_6}{2}\right)}e^{i\eta_{1I}} ,\\
\left(
\begin{array}{c}
A_1^{1+}\\
A_2^{1+}
\end{array}
\right)
&=&
\left(
\begin{array}{c}
e^{\tau_{31}}\\
e^{\tau_{32}}
\end{array}
\right)\frac{e^{-\frac{(R_6+R_7)}{2}}}{(k_1+k_1^*)} .
\eea
\ees

\noindent(b) \underline{\it Soliton 2} ($\eta_{2R} \approx 0$, $\eta_{1R} \rightarrow -\infty$,
$\eta_{3R} \rightarrow \infty$):
\bes
\begin{eqnarray}
\left(
\begin{array}{c}
q_1\\
q_2 
\end{array}
\right)
&\approx&
\left(
\begin{array}{c}
A_1^{2+} \\
A_2^{2+}
\end{array}
\right)k_{2R}
{\mbox{sech}\,\left(\eta_{2R}+\frac{R_6-R_3}{2}\right)}e^{i\eta_{2I}} ,\\
\left(
\begin{array}{c}
A_1^{2+}\\
A_2^{2+}
\end{array}
\right)
&=&
\left(
\begin{array}{c}
e^{\delta_{61}}\\
e^{\delta_{62}}
\end{array}
\right)\frac{e^{-\frac{(R_3+R_6)}{2}}}{(k_2+k_2^*)}.
\eea
\ees

\noindent (c) \underline{\it Soliton 3}  ($\eta_{3R} \approx 0$, $\eta_{1R} \rightarrow -\infty$,
$\eta_{2R} \rightarrow -\infty$):
\bes
\begin{eqnarray}
\left(
\begin{array}{c}
q_1^{3+}\\
q_2^{3+} 
\end{array}
\right)
&\approx &
\left(
\begin{array}{c}
A_1^{3+} \\
A_2^{3+}
\end{array}
\right)k_{3R}
{\mbox{sech}\,\left(\eta_{3R}+\frac{R_3}{2}\right)}e^{i\eta_{3I}} ,\\
\left(
\begin{array}{c}
A_1^{3+} \\
A_2^{3+}
\end{array}
\right)
&=&
\left(
\begin{array}{c}
\alpha_3^{(1)}\\
\alpha_3^{(2)}
\end{array}
\right)\frac{e^{-\frac{R_3}{2}}}{(k_3+k_3^*)}.
\eea
\end{subequations}
The various other quantities are defined below:
\begin{subequations}
\bea
e^{\delta_{1j}}&=&\frac{(k_1-k_2)(\alpha_1^{(j)}\kappa_{21}-\alpha_2^{(j)}\kappa_{11}
)}{(k_1+k_1^*)(k_1^*+k_2)},\;\;
e^{\delta_{6j}}=\frac{(k_2-k_3)(\alpha_2^{(j)}\kappa_{33}-\alpha_3^{(j)}\kappa_{23}
)}{(k_3^*+k_2)(k_3^*+k_3)},\nonumber\;\;j=1,2,\\
e^{\tau_{1j}}&=&\frac{(k_2-k_1)(k_3-k_1)(k_3-k_2)(k_2^*-k_1^*)}
{(k_1^*+k_1)(k_1^*+k_2)(k_1^*+k_3)(k_2^*+k_1)(k_2^*+k_2)(k_2^*+k_3)}\nonumber\\
&&\times
\left[\alpha_1^{(j)}(\kappa_{21}\kappa_{32}-\kappa_{22}\kappa_{31})
+\alpha_2^{(j)}(\kappa_{12}\kappa_{31}-\kappa_{32}\kappa_{11})
+\alpha_3^{(j)}(\kappa_{11}\kappa_{22}-\kappa_{12}\kappa_{21})
\right],\nonumber
\eea
\vspace{-0.7cm}
\bea
e^{\tau_{3j}}&=&\frac{(k_2-k_1)(k_3-k_1)(k_3-k_2)(k_3^*-k_2^*)}
{(k_2^*+k_1)(k_2^*+k_2)(k_2^*+k_3)(k_3^*+k_1)(k_3^*+k_2)(k_3^*+k_3)}\nonumber\\
&&\times
\left[\alpha_1^{(j)}(\kappa_{22}\kappa_{33}-\kappa_{23}\kappa_{32})
+\alpha_2^{(j)}(\kappa_{13}\kappa_{32}-\kappa_{33}\kappa_{12})
+\alpha_3^{(j)}(\kappa_{12}\kappa_{23}-\kappa_{22}\kappa_{13})
\right],\nonumber
\eea
\bea
e^{R_1}=\frac{\kappa_{11}}{k_1+k_1^*},\;\;\;\;\;e^{R_3}=\frac{\kappa_{33}}{k_3+k_3^*},  \nonumber
\eea
\bea
e^{R_4}&=&\frac{(k_2-k_1)(k_2^*-k_1^*)}
{(k_1^*+k_1)(k_1^*+k_2)(k_1+k_2^*)(k_2^*+k_2)}
\left[\kappa_{11}\kappa_{22}-\kappa_{12}\kappa_{21}\right],\nonumber\\
e^{R_6}&=&\frac{(k_3-k_2)(k_3^*-k_2^*)}
{(k_2^*+k_2)(k_2^*+k_3)(k_3^*+k_2)(k_3+k_3^*)}
\left[\kappa_{22}\kappa_{33}-\kappa_{23}\kappa_{32}\right],\nonumber
\eea
\bea
e^{R_7}&=& \frac{|k_1-k_2|^2|k_2-k_3|^2|k_3-k_1|^2}
{(k_1+k_1^*)(k_2+k_2^*)(k_3+k_3^*)|k_1+k_2^*|^2|k_2+k_3^*|^2|k_3+k_1^*|^2}
\nonumber\\
&&\times\left[(\kappa_{11}\kappa_{22}\kappa_{33}-
\kappa_{11}\kappa_{23}\kappa_{32})
+(\kappa_{12}\kappa_{23}\kappa_{31}-
\kappa_{12}\kappa_{21}\kappa_{33})\right .\nonumber\\
&&\left.+(\kappa_{21}\kappa_{13}\kappa_{32}-
\kappa_{22}\kappa_{13}\kappa_{31})\right],\nonumber
\eea
and
\bea
\kappa_{il}= \frac{\sum_{n=1}^2\alpha_i^{(n)}\alpha_l^{(n)*}}
{\left(k_i+k_l^*\right)},\;i,l=1,2,3.\nonumber
\eea
\end{subequations}
\section{Explicit expressions of $A_{1,2}^{2-}$ and $A_{1,2}^{3+}$}
\begin{subequations}
\bea
\left(
\begin{array}{c}
  A_1^{2-} \\
  A_2^{2-}
\end{array}
\right)&&=\frac{1}{\sqrt{|\alpha_2^{(1)}|^2+|\alpha_2^{(2)}|^2}}\bigg(\frac{a_1}{a_1^*}\bigg)\bigg(\frac{\kappa_{12}\kappa_{22}}{\kappa_{21}\kappa_{11}}\bigg)^\frac{1}{2}
\left(
\begin{array}{c}
  \frac{N_1^{2-}}{D^{2-}} \\
 \frac{N_2^{2-}}{D^{2-}}
\end{array}
\right),\\
\left(
\begin{array}{c}
  A_1^{3+} \\
  A_2^{3+}
\end{array}
\right)&&=
\left(
\begin{array}{c}
  \alpha_3^{(1)} \\
  \alpha_3^{(2)}
\end{array}
\right)
\frac{1}{\sqrt{|\alpha_3^{(1)}|^2+|\alpha_3^{(2)}|^2}},
\eea
where 
\bea
N_1^{2-}&&=
\left|
\begin{array}{cc}
\alpha_1^{(1)}&\alpha_2^{(1)} \\
\kappa_{11}&\kappa_{21}
\end{array}
\right|, \quad N_2^{2-}=
\left|
\begin{array}{cc}
\alpha_1^{(2)}&\alpha_2^{(2)} \\
\kappa_{21}&\kappa_{11}
\end{array}
\right|, \quad D^{2-}=
\left|
\begin{array}{cc}
\kappa_{11}&\kappa_{21} \\
\kappa_{12}&\kappa_{22}
\end{array}
\right|^\frac{1}{2},\nonumber\\
a_1&&=(k_1+k_2^*)\bigg[(k_1-k_2)\big(\alpha_1^{(1)*}\alpha_2^{(1)}+\alpha_1^{(2)*} \alpha_2^{(2)}\big)\bigg]^{\frac{1}{2}},\nonumber\\
\kappa_{il}&&=\frac{\alpha_i^{(1)}\alpha_l^{(1)*}+\alpha_i^{(2)}\alpha_l^{(2)*}}{(k_i+k_l^*)}, \quad i,l=1,2.\nonumber
\eea
\end{subequations}
In the above, $*$ represents the complex conjugation.

\end{document}